\documentclass[aps,prb,twocolumn,superscriptaddress]{revtex4}
\usepackage{graphicx}
\usepackage{dcolumn}
\usepackage{amsmath}
\usepackage{amssymb}
\usepackage{subfigure,amsmath,verbatim,moreverb,bm}
\def\be{\begin{equation}}
\def\ee{\end{equation}}
\def\ber{\begin{eqnarray}}
\def\eer{\end{eqnarray}}

\def\rv{{\bf r}}

\def\kv{{\bf k}}
\def\pv{{\bf p}}

\def\Acal{{\bf {\cal A}}}
\begin{document}
\title{Orbital momentum Hall effect in p-doped graphane}
\author{I. V. Tokatly}
\email{Ilya_Tokatly@ehu.es}
\affiliation{Nano-Bio Spectroscopy group and ETSF Scientific Development Centre, 
  Departamento de F\'isica de Materiales, Universidad del Pa\'is Vasco, 
  Centro de F\'isica de Materiales CSIC-UPV/EHU-MPC, E-20018 San Sebasti\'an, Spain}
\affiliation{IKERBASQUE, Basque Foundation for Science, E-48011 Bilbao, Spain}

\date{\today}
\begin{abstract}
It is shown that an electric field applied to p-doped graphane generates a dissipationless orbital momentum Hall current. In the clean limit the corresponding Hall conductivity is independent of the concentration of holes. The Hall effect is related to the $2\pi$-Berry phase accumulated when heavy and light holes are transported around the degeneracy point in the center of the Brillouin zone. This also leads to the orbital momentum edge currents in the equilibrium state, and to the accumulation of the orbital momentum at the edges when the system is driven out of equilibrium. 

\end{abstract}
\maketitle

In the last few years a great deal of attention of the condensed matter community has been attracted by two major developments: an experimental isolation of a high quality single- and multilayer graphene \cite{Novoselov2005,Novoselov2006,Geim2009}, and the discovery of the intrinsic spin Hall effect (SHE) in semiconductor systems with spin-orbit interaction \cite{MurNagZha2003,Sinova2004}. Because of its unique electronic properties \cite{CastroNeto2009} graphene is considered as one of the most prospective materials for future carbon-based electronics \cite{AvoChePer2007}, a possible successor of the standard silicon electronics. SHE is currently regarded as an important part of emergent spintronics \cite{ZutFabSar2004,Fabian2007}.

Recently an important step towards graphene electronics has been made -- an experimental synthesis of a fully hydrogenated graphene, named graphane \cite{Elias2009}. Graphane is a wide band-gap dielectric \cite{SofChaBar2007,Lebegue2009} and therefore it may become an important part of nanoelectronic devices as it opens a way for a controlable creation of a potential landscape with spatial separation of conducting and insulating regions. However, graphane is not only a promising material for passive parts of graphene electronics, but also has very interesting and exciting properties by itself. Theoretical works triggered by the first experiments predict localized spin states at hydrogen vacancies \cite{SahAtaCir2009}, demonstrate the existence of unusual strongly bound charge-transfer excitons \cite{CudAttTok2010arxiv}, and indicate that doped graphane is probably a high-$T_c$ superconductor \cite{SavFerGiu2010arxiv}. 

In this work I show that p-doped graphane demonstrates a dissipationless orbital momentum Hall effect (OHE). In the case of a weak disorder the corresponding Hall conductivity has a value $\sigma^{\rm{oH}}\approx e/\pi\hbar$ independently of the density of holes. Similarly to the intrinsic SHE in systems with spin-orbit interaction \cite{MurNagZha2003,Sinova2004}, the origin of OHE in graphane is a Berry phase related to an effective ${\bf k}$-space ``magnetic monopole'' (Aharonov-Bohm flux line in this case) located at the band degeneracy point. In the equilibrium state it generates dissipationless orbital momentum edge currents, while in the presence of a transport charge current it leads to the accumulation of the orbital momentum at the edges of the sample. These findings make a direct connection of graphane to the field of spintronics. In fact, they suggest a potential application of graphane as an active material for spintronics, or, more precisely, orbitronics \cite{BerHugZha2005} devices, which can be naturally integrated within future graphene-based electronics chips.

Graphane in its most stable chair conformation \cite{SofChaBar2007,Lebegue2009} is obtained from the ideal graphene by depositing hydrogen on both sides of the graphene plane and hydrogenating the A- and B-sublattice carbons from the upper and the lower side, respectively. The A (B) carbons relax upwards (downwards) and form $\sigma$-bonds with the corresponding H-atoms, which results in a diamond-like $sp^3$ hybridization of carbon orbitals. This has a dramatic effect on the electronic structure. Graphane is a wide, direct band-gap insulator with the gap minimum of 5.4~eV at the $\Gamma$-point (center of the Brillouin zone) \cite{SofChaBar2007,Lebegue2009}. 

States at the top of the valence band, which are mainly made of the carbon $p$-orbitals, belong to a two-dimensional representation $E_{g}$ of the graphane point group $D_{3d}$ \cite{CudAttTok2010arxiv}. Hence the corresponding $k\cdot{p}$ Hamiltonian should be a 2$\times$2 matrix \cite{spin-orbit}. In general the symmetry fixes the $k\cdot{p}$ matrix up to a unitary transformation (choice of basis functions of the representation). In Ref.~\onlinecite{CudAttTok2010arxiv} an effective Hamiltonian for holes in graphane has been derived using a set of $\{x^2-y^2,2xy\}$ as a basis of $E_g$ representation. In the present context a unitary equivalent form of the Hamiltonian, which corresponds to the basis $\{(x+iy)^2,(x-iy)^2\}$, is more natural:
\begin{eqnarray}
 \nonumber
\hat{H}&=&\frac{1}{2}\gamma_1I\hat{\pv}^2  +\frac{1}{4}\gamma_2[\sigma_{+}\hat{p}_{+}^2+\sigma_{-}\hat{p}_{-}^2]\\
&\equiv& \frac{1}{2}
\left[\begin{array}{cc}
    \gamma_1(\hat{p}_x^2+\hat{p}_y^2)  &  \gamma_2(\hat{p}_{x}+i\hat{p}_{y})^2\\
 \gamma_2(\hat{p}_{x}-i\hat{p}_{y})^2  &  \gamma_1(\hat{p}_x^2+\hat{p}_y^2)
\end{array}\right]
\label{kpH}
\end{eqnarray}
where $\hat{\pv}=-i\nabla$, $\hat{p}_{\pm}=\hat{p}_{x}\pm i\hat{p}_{y}$, $I$ is the identity matrix, and $\sigma_{\pm}=\sigma_{x}\pm i\sigma_{y}$ with $\sigma_j$ being the Pauli matrices. The constants $\gamma_1>\gamma_2$ found in \cite{CudAttTok2010arxiv} determine the effective masses of the isotropic light (L) and heavy (H) hole bands, $m_{L}=(\gamma_1+\gamma_2)^{-1}\approx 0.22m_0$ and $m_{H}=(\gamma_1-\gamma_2)^{-1}\approx 0.64m_0$, where $m_0$ is the bare electronic mass. 

\begin{figure}
\begin{center}
\includegraphics[width=0.7\linewidth]{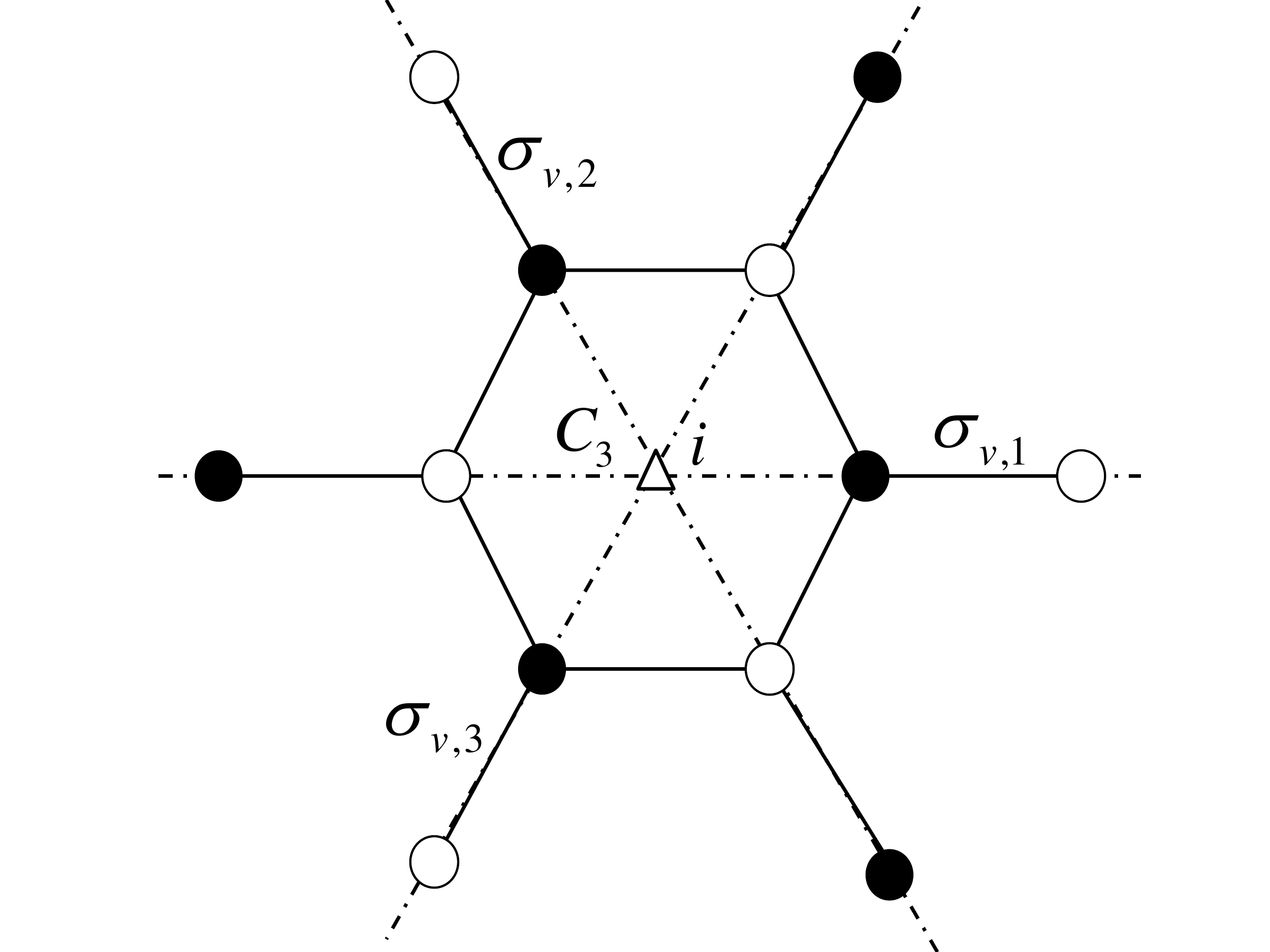}
\caption{Schematic structure of graphane plane and the symmetry elements of $D_{3d}$ point group: the center of inversion $i$ and the third order axis $C_3$ at the center of the hexagon, and three vertical reflection planes $\sigma_{v,1}$, $\sigma_{v,2}$ and $\sigma_{v,3}$. \label{fig:one}}
\end{center}
\end{figure}

It is instructive to verify explicitly that $\hat{H}$ of Eq.~\ref{kpH} is indeed the proper invariant. Operations of the group $D_{3d}\sim C_{i}\times C_{3v}$ are shown on Fig.~1. These are the inversion $i$, in-plane rotations by $2\pi/3$ about $C_3$ axis, and reflections in three vertical planes  $\sigma_{v,1}$, $\sigma_{v,2}$, and $\sigma_{v,3}$. Acting by the group operations on functions $\phi_{\pm}(x,y)=(x\pm iy)^2$ one constructs a 2$\times$2 matrix representation of the group in this basis \cite{reflections}: $\hat{D}(i)=I$,
\begin{equation}
 \label{operations}
\hat{D}(C_{3})=e^{i\frac{2\pi}{3}\sigma_z}, \quad{\rm and}\quad \hat{D}(\sigma_{v,1})=\sigma_x.
\end{equation}
Using Eq.~(\ref{operations}) we immediately find that matrices $\sigma_{+}$ and $\sigma_{-}$ belong to the representation $E_g$ and transform according to the rules, $\sigma_{\pm}\xrightarrow{C_3}e^{\pm i 4\pi/3}\sigma_{\pm}$ and $\sigma_{\pm}\xrightarrow{\sigma_{v,1}}\sigma_{\mp}$, which are identical to those for the operators $\hat{p}_{\pm}^2$. Hence the term $\sigma_{+}\hat{p}_{+}^2+\sigma_{-}\hat{p}_{-}^2$ in Eq.~(\ref{kpH}) is the group invariant. Similarly the second invariant entering Eq.~(\ref{kpH}) is a product of the identity matrix $I$ and the operator $\hat{\pv}^2$, which both belong to the one dimensional representation $A_{1g}$.
The remaining linearly independent 2$\times$2 matrix $\sigma_z$ is rotationally and inversion invariant, but antisymmetric under reflections. Hence it can enter the Hamiltonian only in the presence of a magnetic field ${\bf B}$, in a form of $\sigma_{z}B_{z}$. It is worth noting that the Hamiltonian of Eq.~(\ref{kpH}) is time reversal invariant with the operator of time inversion $\hat{\Theta}=\sigma_{x}K$, where $K$ stands for the complex conjugation. Interestingly, for $\gamma_1=0$ Eq.~(\ref{kpH}) coinsides with the Hamiltonian of bilayer graphene \cite{McCFal2006,Novoselov2006}.

Since matrix $\sigma_z$ acts a generator of rotations [see Eq.~(\ref{operations})], and couples to the magnetic field as $\sigma_{z}B_{z}$, it is naturally identified with the $z$-component of the operator of a local orbital momentum, i.~e., the angular momentum of orbital currents inside the unit cell of graphane. This identification is vital for all subsequent discussion.

Eigenstates of the Hamiltonian Eq.~(\ref{kpH}) correspond to two parabolic bands with dispersions $E_{k}^{L,H}={k^2}/{2m_{L,H}}$ and the plane-wave spinor wave functions of the form $\Psi_{L,H}=\Phi^{L,H}_{\kv} e^{i\kv\rv}$, where
\begin{equation}
 \label{wave-fun}
\Phi^{L}_{k_x,k_y}=\frac{1}{\sqrt{2}}\left(\begin{array}{r}
                         \hat{k}_{+}\\
			\hat{k}_{-}
                        \end{array}\right), 
\quad
\Phi^{H}_{k_x,k_y}=\frac{1}{\sqrt{2}}\left(\begin{array}{r}
                         \hat{k}_{+}\\
			-\hat{k}_{-}
                        \end{array}\right),
\end{equation}
and $\hat{k}_{\pm}=(k_x\pm ik_{y})/k$ (here $k=|\kv|$). It is customary to describe states of Eq.~(\ref{wave-fun}) as chiral \cite{Novoselov2006} because they are eigenfunctions (with eigenvalues $\pm 1$) of a ``chirality'' operator $\bm\sigma{\bf n}_{\kv}$, where ${\bf n}_{\kv}=(\cos 2\theta_{\kv},-\sin 2\theta_{\kv})$, and $\theta_{\kv}$ is a polar angle of the wave vector $\kv$. To uncover the physical significance of the chirality for graphane (as well as for bilayer graphene) we note that the chirality operator can be represented as follows
\begin{equation}
 \label{chirality}
\bm\sigma{\bf n}_{\kv}= e^{i\theta_{\kv}\sigma_z}\sigma_{x}e^{-i\theta_{\kv}\sigma_z}\equiv\hat{D}(\sigma_{v,\kv}),
\end{equation}
which is exactly the operator of reflection in a vertical plane $\sigma_{v,\kv}$ parallel to the wave vector. Hence the chirality of the L and H holes is identical to a parity, positive and negative, respectively, under reflection in the ``propagation plane''. We will make use of this fact later on.

An important topological property of the L and H states, Eq.~(\ref{wave-fun}), is a nontrivial Berry connection
\begin{equation}
 \label{connection}
\Acal_{\kv}=i\left\langle\Phi^{L}_{\kv}|\nabla_{\kv}\Phi^{H}_{\kv}\right\rangle =\frac{\kv\times\hat{\bf z}}{k^2},
\end{equation}
which is related to the band degeneracy at the point $\kv=0$. In fact, the degeneracy point plays a role of an effective Aharonov-Bohm magnetic line which carries a flux (Berry phase) of $2\pi$
\begin{equation}
 \label{flux}
\oint\Acal_{\kv} d{\bf l}_{\kv}=2\pi
\end{equation}

Now we are ready to discuss OHE in graphane. Since the operator of the local orbital momentum is identified as $\hat{L}_{\rm{loc}}^z=\sigma_z$ we can define an operator of the orbital momentum current as follows
\begin{equation}
 \label{J-operator}
\hat{\bf J}^z = \frac{1}{2}\{\sigma_z,\hat{\bf V}\},
\end{equation}
where $\hat{\bf V}=\nabla_{\kv}\hat{H}$ is the velocity operator, and $\{A,B\}=AB+BA$. 

The operator of Eq.~(\ref{J-operator}) is odd under reflection in the $\kv$-plane. Since the states of Eq.~(\ref{wave-fun}) are the eigenfunctions of this operation, all diagonal matrix elements of $\hat{\bf J}^z$ are zero, which implies a vanishing bulk orbital momentum current in the equilibrium. However a nontrivial Berry flux suggests that a nonzero transverse current ${\bf J}^z$ can be generated by applying an external electric field ${\bf E}e^{i\omega t}$. Using the standard linear response theory we get
\begin{equation}
 \label{J-resp}
J_i^z = \sigma_{ij}^{\rm{oH}}(\omega)E_j
\end{equation}
\begin{equation}
 \label{Kubo}
\sigma_{ij}^{\rm{oH}}(\omega) =\frac{2i}{\omega}\sum_{\kv}\sum_{\alpha,\beta=L,H}(f_{k}^{\alpha}-f_{k}^{\beta})
\frac{\langle\alpha|\hat{J}_i^z|\beta\rangle\langle\beta|\hat{V}_j|\alpha\rangle}{\omega - (E_{k}^{\beta}
-E_{k}^{\alpha})}
\end{equation}
where $f_{k}^{\alpha}=f(E_{k}^{\alpha})$ is the Fermi distribution function. After straightforward algebra (the reflection symmetry ensures that only off-diagonal matrix elements contribute) Eq.~(\ref{Kubo}) simplifies as follows, $\sigma_{ij}^{\rm{oH}}(\omega)=\varepsilon_{ij}\sigma^{\rm{oH}}_{\omega}$, where
\begin{equation}
 \label{sigma1}
\sigma^{\rm{oH}}_{\omega}=\sum_{\kv}(f_{k}^{H}-f_{k}^{L})
\frac{[\Acal_{\kv}\times\nabla_{\kv}(E_k^L+E_k^H)]\Delta(k)}{\Delta^2(k)-\omega^2},
\end{equation}
and $\Delta(k)=E_k^L - E_k^H=k^2/2m_L-k^2/2m_H$ is the energy gap between L and H bands at a given $k$. Since for parabolic bands $E_k^L+E_k^H\sim\Delta(k)$, the last factor in Eq.~(\ref{sigma1}) is proportional to the cross product of the Berry connection and a total derivative, $\Acal_{\kv}\times\nabla_{\kv}\ln [\Delta^2(k)-\omega^2]$. Thus at $T\to 0$ the right hand side of Eq.~(\ref{sigma1}) reduces (by partial integration) to a Fermi contour integral that yields, as expected, the Hall conductivity proportional to the Berry flux of Eq.~(\ref{flux}). The final result takes the form
\begin{equation}
 \label{sigma2}
\sigma^{\rm{oH}}_{\omega}=\frac{1}{4\pi}\frac{m_H+m_L}{m_H-m_L}\ln
\left[\frac{\Delta^2(k_F^H)-\omega^2}{\Delta^2(k_F^L)-\omega^2}\right],
\end{equation}
where $k_F^H=\sqrt{2m_H\varepsilon_F}$ and $k_F^L=\sqrt{2m_L\varepsilon_F}$ are the Fermi momenta of the H and L holes, respectively. In the limit $\omega\to 0$ the Fermi energy in Eq.~(\ref{sigma2}) cancels out and we get the following static Hall conductivity
\begin{equation}
 \label{sigma0}
\sigma^{\rm{oH}}_0 = \frac{1}{2\pi}\frac{m_H+m_L}{m_H-m_L}\ln\left(\frac{m_H}{m_L}\right)
\end{equation}
independently of the concentration of holes \cite{concentration}. Using the actual masses and recovering physical units we obtain the value $\sigma^{\rm{oH}}_0\approx e/\pi\hbar$, which closely reminds the ``universal'' intrinsic spin Hall conductivity predicted for the 2D electron gas with spin-orbit interaction \cite{Sinova2004}. The crucial difference is that the result of Eq.~(\ref{sigma0}) is insensitive to a weak disorder -- vertex corrections vanish because the Hamiltonian (\ref{kpH}) is quadratic in the momentum \cite{Murakami2004}. The final static Hall conductivity in the presence of a $\delta$-correlated disorder is obtained from Eq.~(\ref{sigma2}) by replacing $\omega\to i/\tau$, where $\tau$ is the momentum relaxation time. 

Recently a nonzero orbital momentum Hall conductivity has been predicted for p-doped Si \cite{BerHugZha2005}. It is instructive to compare OHE in graphane and in Si with intrinsic SHE \cite{MurNagZha2003,Sinova2004}. While OHE in graphane is an analog of the intrinsic SHE in the 2D Rashba system \cite{Sinova2004}, the OHE in Si is closely related to SHE of holes in 3D semiconductors \cite{MurNagZha2003}. These similarities are related to similar gauge structures underlying OHE and SHE in the above systems.

A symmetry reason for the dissipationless static OHE is that the external electric field breaks the reflection symmetry in the plane perpendicular to the direction of ${\bf E}$. This allows for a nonzero orbital momentum current propagating along that plane, i.~e. OHE. Similarly one would expect that even in the equilibrium any reflection symmetry breaking defect, e.~g. the boundary, will generate orbital momentum currents flowing along the defect's potential isolines. Below I explicitly demonstrate the existence of such equilibrium edge currents for a plane boundary of graphane, modelled by an infinite potential wall along the line $y=0$. In other words, we consider the ``edge defect'' with the potential $V(\rv)=0$ at $y>0$, and $V(\rv)=\infty$ at $y<0$.

In the presence of an infinite, ideally reflecting potential wall the orthonormal set of scattering states labelled by the incident momentum $\kv=(k_x,-k_y)$, $k_y>0$ takes the form
\begin{eqnarray}
 \nonumber
\Psi_{\kv}^L(\rv) &=& \Big[\Phi_{k_x,-k_y}^Le^{-ik_yy} - R_{\kv}^{L}\Phi_{k_x,k_y}^Le^{ik_yy} \\ 
                  &+& C_{\kv}^{L}\Phi_{k_x,p_H}^He^{ip_Hy}\Big]e^{ik_xx},            \label{Lscatt}
\\ \nonumber
\Psi_{\kv}^H(\rv) &=& \Big[\Phi_{k_x,-k_y}^He^{-ik_yy} - R_{\kv}^{H}\Phi_{k_x,k_y}^He^{ik_yy} \\ 
                  &+& C_{\kv}^H\Phi_{k_x,p_L}^Le^{ip_Ly}\Big]e^{ik_xx},             \label{Hscatt}
\end{eqnarray}
where the spinors $\Phi_{k,p}^{L}$ and $\Phi_{k,p}^{L}$ are defined by Eq.~(\ref{wave-fun}). The kinematics of scattering corresponding to the solutions of Eqs.~(\ref{Lscatt}) and (\ref{Hscatt}) is illustrated on Fig.~2. 
\begin{figure}
\begin{center}
\includegraphics[width=0.9\linewidth]{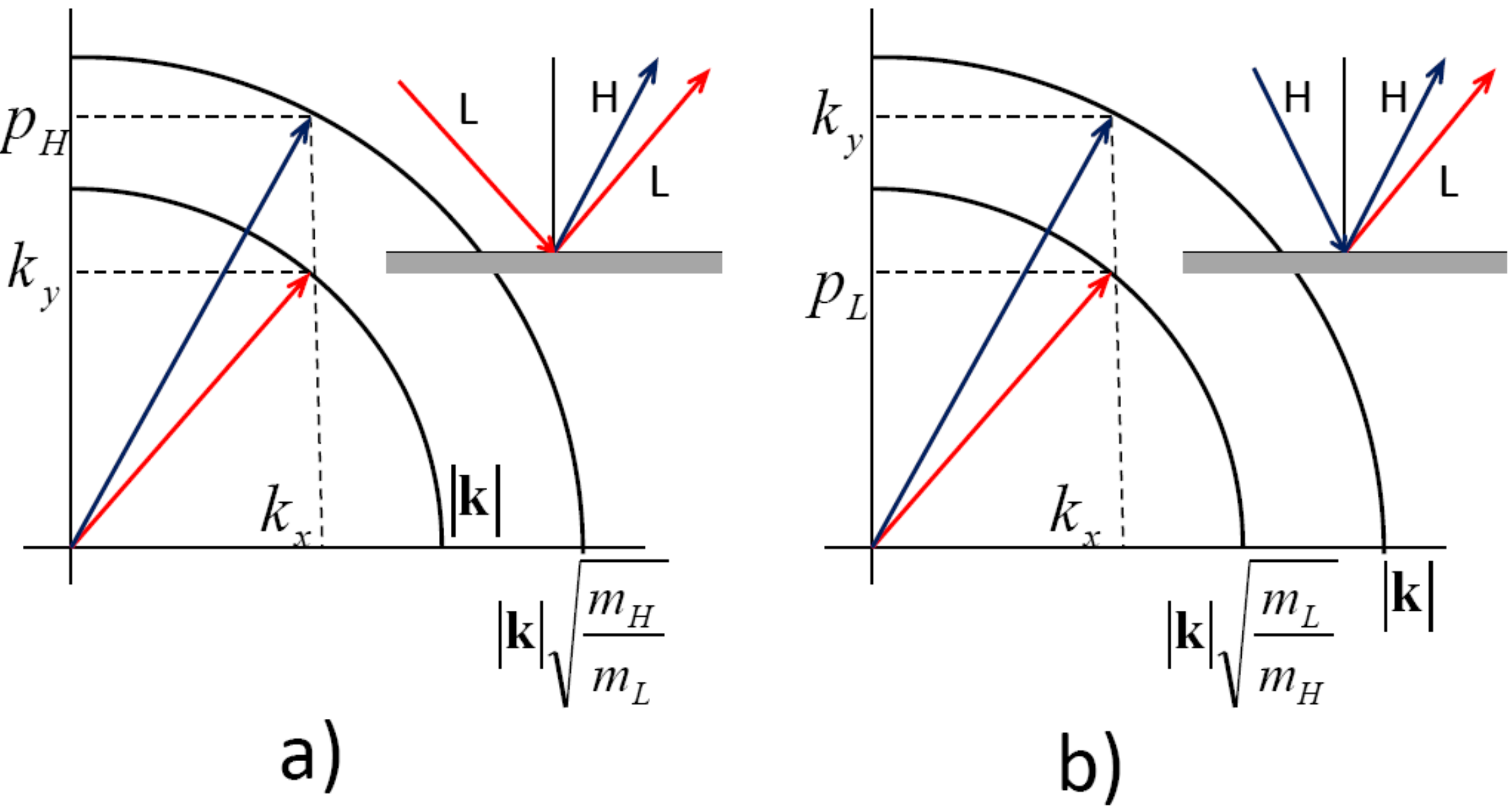}
\caption{(Color online) Geometry of scattering and mixing of the light (L) and heavy (H) holes at the edge of graphane. \label{fig:two}}
\end{center}
\end{figure}
The states $\Psi_{\kv}^L(\rv)$ and $\Psi_{\kv}^H(\rv)$ originate from incident L and H waves, the first terms in Eqs.~(\ref{Lscatt}) and (\ref{Hscatt}). The second terms are the reflected L (H) waves with the reflection coefficients $R_{\kv}^{L(H)}$. Importantly, the incident L (H) holes also produce reflected H (L) states, the third terms in Eqs.~(\ref{Lscatt}) and (\ref{Hscatt}), with amplitudes given by L$\to$H (H$\to$L) ``conversion'' coefficients $C_{\kv}^{L(H)}$. The $y$-components $p_H$ ($p_L$) of the momentum of the L$\to$H (H$\to$L) reflected waves  are fixed by kinematic constraints of the conservation of energy and the $x$-component of momentum (see Fig.~2): 
$$
p_H=\sqrt{\frac{m_H}{m_L}k^2-k_x^2}, \quad p_L=\sqrt{\frac{m_L}{m_H}k^2-k_x^2}.
$$
Finally, the reflection and conversion coefficients are determined from the infinite wall boundary conditions, $\Psi_{\kv}^{L,H}|_{y=0}=0$. For the incident H-hole we find:
\begin{equation}
\label{reflection}
R_{\kv}^{H}=\frac{k_{x}^2 - k_{y}p_{L}}{k_{x}^2 + k_{y}p_{L}}, \quad
C_{\kv}^{H}=i\sqrt{\frac{m_L}{m_H}}\frac{2k_{x}k_{y}}{k_{x}^2 + k_{y}p_{L}}
\end{equation}
(the results for the incident L-hole are obtained by interchanging L and H indexes in this equation). Note that there is a range of incidence angles for the H-hole when the $y$-component $p_L$ of the momentum of the reflected L-hole is purely imaginary, i.~e., the corresponding L-state becomes an evanescent wave localized near the boundary.

The L-H mixing in the scattering states Eqs.~(\ref{Lscatt}), (\ref{Hscatt}) is related to the broken reflection symmetry. We can see from Eq.~(\ref{reflection}) that the conversion coefficients vanish only for the normal incidence, $k_x=0$, as only in this case the hole states are still the eigenfunctions of the reflection/chirality operator, Eq.~(\ref{chirality}).

The equilibrium orbital momentum current flowing along the edge is given by the expression
\begin{equation}
 \label{edge-curent_def}
J_{x}^{z}(y) = 2\sum_{\alpha=L,H}{\sum_{\kv}}' f_{k}^{\alpha}[\Psi_{\kv}^{\alpha}(\rv)]^{\dagger}\hat{J}_{x}^z\Psi_{\kv}^{\alpha}(\rv),
\end{equation}
where the prime means that the summation is restricted to $k_y>0$. Explicitly Eq.~(\ref{edge-curent_def}) can be represented in the following compact form
\begin{equation}
 \label{edge-current1}
J_{x}^{z}(y) =\frac{2}{m_{r}}{\rm Im}{\sum_{\kv}}'f_k^{H}\frac{k_x^2k_y}{k^2}
R_{\kv}^H\left(e^{ik_yy} - e^{ip_Ly}\right)^2
\end{equation}
where $m_r$ is a reduced mass, $m_r^{-1} = m_H^{-1}+m_L^{-1}$. From the structure of Eq.~(\ref{edge-current1}) it is clear that $J_{x}^{z}(y)$ decays away from the edge and experiences Friedel oscillations.

For the total equilibrium edge current at $T\ll\varepsilon_F$ we get the following simple result
\begin{equation}
 \label{edge-current2}
\bar{J}_x^z=\int_{0}^{\infty}J_{x}^{z}(y)dy = \varepsilon_{F}\frac{\pi (m_H - m_L)^2}{8m_Hm_L}.
\end{equation}
Hence the net edge current is proportional to the Fermi energy -- the result expected on dimensional grounds as $\varepsilon_F$ is the only parameter of the proper dimension [$\hbar/t$].

Now I demonsrate that OHE leads to the orbital momentum accumulation at the edges of a graphane sample. The local orbital momentum operator $\sigma_z$ is antisymmetric under the time reversal and the reflections. Therefore both symmetries must be broken to have a nonzero expectation value $\langle\sigma_z\rangle$. Let us first analyze the bulk situation. Because of the reflection symmetry the orbital momentum vanishes independently for every plane wave state of Eq.~(\ref{wave-fun}), which in particular implies $\langle\sigma_z\rangle=0$ in the equilibrium. Breaking the time reversal symmetry and one of the reflections by applying an external electric field $E$ and generating a transport charge current $j$ still does not produce a nonzero $\langle\sigma_z\rangle$. The reason is the remaining reflection symmetry in the vertical plane parallel to $j$. The presence of the edge breaks that symmetry and thus allows for a nonzero orbital momentum, which should be localized at the edge as it must vanish in the bulk of graphane. This symmetry argumentation can be explicitly illustrated for the model boundary considered above. Because of the broken reflection symmetry the orbital momentum $[\Psi_{\kv}^{\alpha}(\rv)]^{\dagger}\sigma_z\Psi_{\kv}^{\alpha}(\rv)$ is nonzero for all scattering states, except for the normal incidence. However, it is proportional to $k_x$ (as dictated by the time reversal symmetry) and thus vanishes when integrated with any symmetric in $k_x$ distribution function. In the transport situation the distribution function becomes asymmetric and the integration yields a nonzero orbital momentum localized at the edge. On dimensional grounds we conclude that the line density of the edge orbital momentum should be proportional to $\hbar^2j/\varepsilon_F\sim e\tau E$, which is the momentum asymmetry responsible for the transport current. This is similar to the spin accumulation at the surface of semiconductors described by the Luttinger Hamiltonian \cite{StaGal2006}.

In conclusion I predict OHE for p-doped graphane with all standard features of the Hall effect present: a nonzero dissipationless Hall conductivity, the equilibrium edge currents (which are reminiscent of the Landau diamagnetic currents), and accumulation of the orbital momentum at the edges. The later effect should be ovservable experimentally via detecting the current induced magnetization of the opposite sign at the opposite edges of the graphane sample. 

This work was supported by
the Spanish MEC (FIS2007-65702-C02-01), ``Grupos Consolidados UPV/EHU del Gobierno
Vasco'' (IT-319-07), and the European Union through e-I3 ETSF project (Contract No. 211956).


\begin{thebibliography}{30}
\bibitem{spin-orbit}
\bibinfo{note}{We neglect a weak spin-orbit interaction in graphane.}
\bibitem{reflections}
\bibinfo{note}{Since all reflections belong to the same class it is sufficient to
  consider explicitly only $\sigma_{v,1}$.}
\bibitem{concentration}
\bibinfo{note}{Aparently inclusion of interactions and/or disorder will renormalize this result and introduce a weak density dependence.}
\end{thebibliography}

\end{document}